\documentclass{aa}
\usepackage{graphicx}
\usepackage{txfonts}
\usepackage{orcidlink}

\begin{document} 

   \title{Magnetohydrodynamic simulation assessment of a potential near-ultraviolet early ingress in WASP-189b}

   \author{Y. Duann
          \orcidlink{0000-0002-4260-385X}
          \inst{1,2,3}\fnmsep\thanks{Corresponding Author: yi.duann@sund.ku.dk}
          \and
          S.-H. Lai
          \orcidlink{0000-0002-9024-3326}
          \inst{2,3}
          \and
          H. J. Hoeijmakers
          \orcidlink{0000-0001-8981-6759}
          \inst{4}
          \and
          A. Johansen
          \orcidlink{0000-0002-5893-6165}
          \inst{1,4}
          \and
          C.-L. Lin
          \orcidlink{0000-0001-5989-7594}
          \inst{5,6}
          \and
          L.-C. Huang
          \orcidlink{0000-0002-9679-5279}
          \inst{7}
          \and\newline
          Y.-Y. Chang
          \orcidlink{0009-0008-9937-0330}
          \inst{2,3}
          \and
          A. G. Sreejith
          \orcidlink{0000-0002-4166-4263}
          \inst{8,9}
          \and
          K. France
          \orcidlink{0000-0002-1002-3674}
          \inst{8}
          \and
          L. C. Chang
          \orcidlink{0000-0002-6495-1185}
          \inst{2,3}\fnmsep\thanks{Corresponding Author: loren@g.ncu.edu.tw}
          \and
          W.-H. Ip
          \orcidlink{0000-0002-3140-5014}
          \inst{2,5}
          }
    \titlerunning{MHD simulation assessment of a potential NUV early ingress in WASP-189b}
    \authorrunning{Y. Duann et al.}

   \institute{Globe Institute--Center for Star and Planet Formation, University of Copenhagen, Copenhagen, Denmark
        \and
        Department of Space Science and Engineering, National Central University, Taoyuan, Taiwan
        \and
        Center for Astronautical Physics and Engineering, National Central University, Taoyuan, Taiwan
        \and
        Department of Physics--Lund Observatory, Lund University, Lund, Sweden
        \and
        Institute of Astronomy, National Central University, Taoyuan, Taiwan
        \and
        Steward Observatory, The University of Arizona, Tucson, AZ, USA
        \and
        Shanghai Astronomical Observatory, Chinese Academy of Sciences, Shanghai, China
        \and
        Laboratory for Atmospheric and Space Physics, University of Colorado Boulder, Boulder, CO, USA
        \and
        Space Research Institute, Austrian Academy of Sciences, Graz, Austria
             }

   \date{Received 1 July 2025/ Accepted 17 September 2025}

\abstract 
{Ultra-hot Jupiters (UHJs) in close orbits around early-type stars provide natural laboratories for studying atmospheric escape and star–planet interactions under extreme irradiation and wind conditions. The near-ultraviolet (NUV) regime is particularly sensitive to extended upper atmospheric and magnetospheric structures.}
{We investigate whether star–planet interactions in the WASP-189 system could plausibly account for the early ingress feature suggested by NUV transit fitting models.}
{We analysed three NUV transits of WASP-189b observed as part of the Colorado Ultraviolet Transit Experiment (CUTE), which employs a 6U CubeSat dedicated to exoplanet spectroscopy. To explore whether the observed transit asymmetry could plausibly arise from a magnetospheric bow shock (MBS), we performed magnetohydrodynamic (MHD) simulations using representative stellar wind velocities and planetary atmospheric densities.}
{During Visit~3, we identified a $\sim$$31.5$-minute phase offset that is consistent with an early ingress. Our MHD simulations indicate that, with a wind speed of $572.97$~$\rm km~s^{-1}$ and a sufficient upper atmospheric density ($\sim$$4.59\times10^{-11}~\rm kg~m^{-3}$), a higher-density zone due to compression can form ahead of the planet within five planetary radii in regions where the fast-mode Mach number falls below $\sim$0.56, even without a MBS. Shock cooling and crossing time estimates from the simulations further suggest that such a pileup could, in principle, produce detectable NUV absorption.}
{Our results indicate that while MBS formation is feasible for WASP-189b, low stellar-wind speeds favour NUV-detectable magnetic pileups over classical bow shocks. Immediately after the shock formation, the post-shock plasma is too hot for strong NUV absorption, but a high-to-low wind-speed transition shortens the cooling time while preserving the compressed plasma, increasing its opacity. Pressure-balance estimates show that magnetic pressure dominates across wind regimes in the low-density case, and at low wind speeds in the high-density case, favouring pileup and reconnection near the magnetopause and enhancing the potential detectability of early-ingress signatures.}


   \maketitle

\section{Introduction}\label{sec1}

Ultra-hot Jupiters (UHJs), orbiting close to luminous stars, are exposed to extreme irradiation, which drives atmospheric expansion, rapid mass loss, and metal-enriched exospheres~\citep{Parmentier2018,Helling2019,Arcangeli2018,Lothringer2018,Hoeijmakers2019}. These systems serve as natural laboratories for studying atmospheric dynamics and star–planet interactions, particularly via near-ultraviolet (NUV) transmission spectroscopy~\citep{Lai2010,Llama2011,Vidotto2011,Alexander2015,Stangret2021}. NUV transit anomalies, such as the early ingress (EI) reported for WASP-12b, have been interpreted as possible signatures of planetary magnetospheric bow shocks \citep[MBSs;][]{Vidotto2010,Vidotto2011,Llama2011,Haswell2012,Wong2022}, with simulations highlighting the key role of magnetic topology in shaping stellar wind–planet interactions~\citep{Debrecht2018,Daley2019}. However, the MBS scenario remains debated, as models indicate that shock cooling timescales may be too long to produce sufficient NUV opacity~\citep{Alexander2015}.

WASP-189b, a UHJ orbiting the A-type star HD 133112 every $\sim$$2.72$ days, exhibits extreme dayside temperatures ($>3400$ K), a highly inclined orbit, and pronounced stellar gravity darkening~\citep{Lendl2020,Deline2022,Yan2020}. Observations reveal an extended, metal-rich upper atmosphere and signatures suggestive of complex magnetospheric interactions~\citep{Prinoth2022,Prinoth2023,Sreejith2023}. Motivated by the fitted NUV EI in this system, we investigated whether such features can plausibly arise from star–planet interaction processes, such as MBS formation, or due to zones of higher density, and discuss their implications for atmospheric escape and magnetospheric dynamics in UHJs.

\section{Observations and data analysis}

\subsection{CUTE NUV transit observations and fittings}

WASP-189b was observed in the NUV (2479–3306~\AA, $R\sim$$750$~W) by the Colorado Ultraviolet Transit Experiment (CUTE) 6U CubeSat, which enables consecutive transit coverage thanks to its 96-minute orbit \citep{Sreejith2023,France2023}. We analysed the same three-transit dataset as \citet{Sreejith2022,Sreejith2023}, processed with their pipeline, excluding exposures with jitter $>$6$''$ or charge-coupled device (CCD) temperatures $>-5^\circ$C. V1 includes 13 orbits; V2 and V3 include 16 each, over phases $-0.2$ to $+0.2$.

White-light curves were constructed by integrating all wavelengths and fluxes reported in detector counts. Systematics were corrected with polynomial fits and robust transit parameters obtained via Markov chain Monte Carlo (MCMC) analysis. Transmission spectra were extracted using the `divide-white' method \citep{Kreidberg2014}. Due to limited S/N, spectral anomalies could not be attributed to specific species.

\begin{figure}[hbt!]
    \centering
    \includegraphics[width=0.5\textwidth]{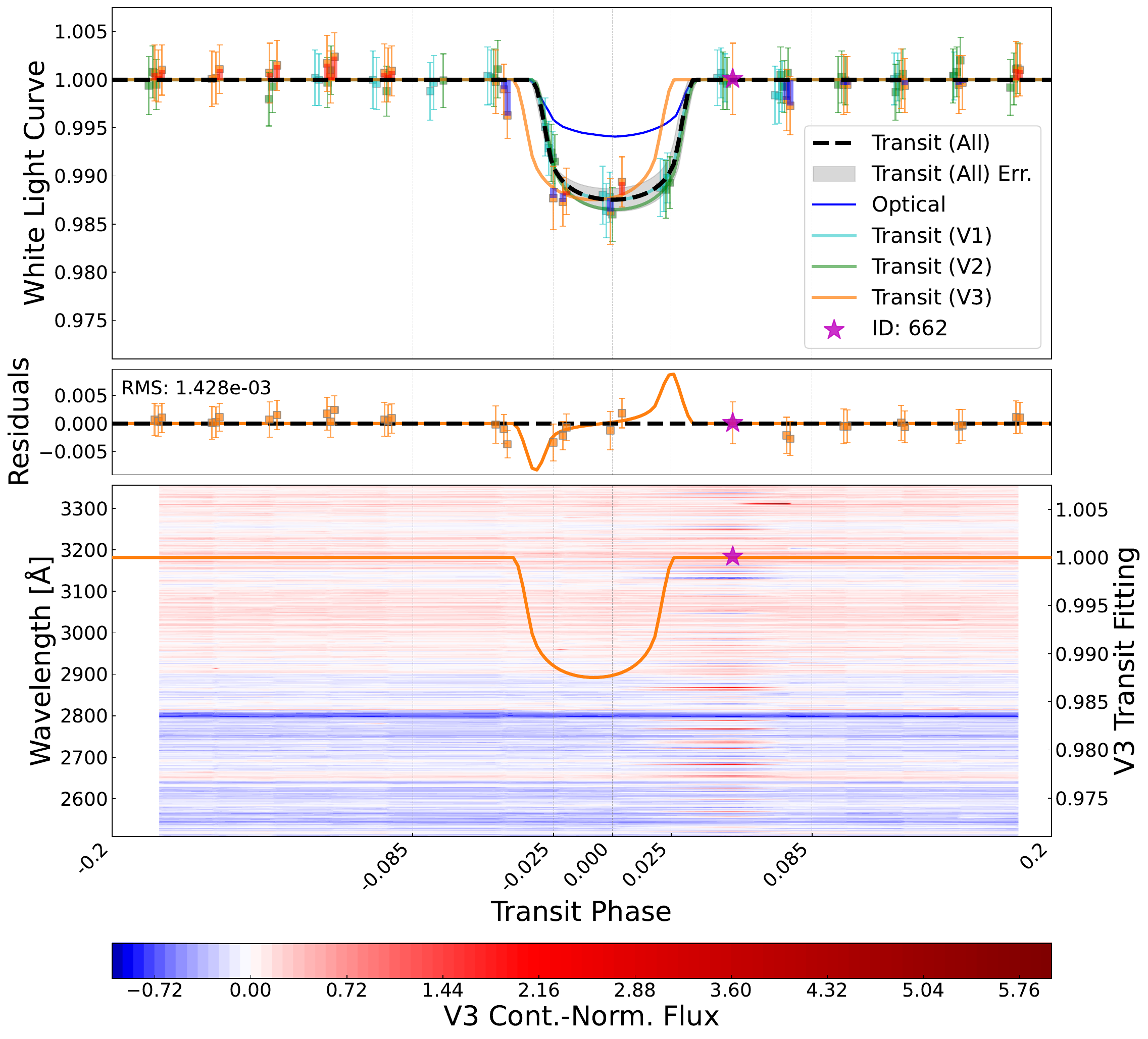}
    \caption{CUTE NUV transit and spectral analysis for WASP-189b.
    Top: White-light curves for three visits (V1, V2, and V3; coloured points) and all data combined, with best-fit transit models (dashed lines) and uncertainties (grey shading). The magenta star marks Event 662 in V3.
    Middle: Residuals relative to best-fit models for each visit and all data, highlighting deviations in V3 (orange).
    Bottom: Spectral map for V3: flux as a function of wavelength and phase, normalised to the continuum/ The colour scale shows fractional deviations. Anomalous features after egress (Event 662) indicate transient variability.}\label{fig1}
\end{figure}

Transit fitting with \texttt{batman} and \texttt{emcee} \citep{Kreidberg2015,ForemanMackey2013}, using CHEOPS parameters \citep{Lendl2020} and PHOENIX limb-darkening, produced generally good fits (Table~\ref{tab:wasp189b_parameters}). For V3, a $\sim$31.5-minute ingress offset was found (Fig.~\ref{fig1}), corresponding to a possible absorbing extension to $\sim$3.65~$R_\mathrm{p}$ based on standard geometric calculations \citep{Anderson2018}. The RMS residual ($\sim$$10^{-3}$) supports the model quality, but the physical origin of this offset remains unknown. In this work we assessed whether a MBS could produce the EI signature in WASP-189b, treating the measured offset as an upper bound. Our Bayesian analysis (Appendix~\ref{EI_Bayes}) yields moderate evidence of an EI feature in Visit 3 (V3). A uniform phase correction was applied across all transits for consistency, and limited phase coverage renders this feature tentative; related Mg~\textsc{ii} spectral variations are shown in Fig.~\ref{append_mgii}.

\begin{table}
   \caption{Input parameters and best-fit results from the MCMC analysis.}\label{tab:wasp189b_parameters}
   $$
   \begin{array}{llll}
      \hline
      \noalign{\smallskip}
      \multicolumn{4}{c}{\text{WASP-189b Planetary Parameters}} \\
      \hline
      \text{Symbol} & \text{Unit} & \text{Value} \\
      \hline
      P & \text{days} & 2.7240330~\tablefootmark{a} \\
      a/R_\mathrm{s} & -- & 4.67~\tablefootmark{a,b} \\
      i & \text{deg} & 84.03~\tablefootmark{b} \\
      e & -- & 0~\tablefootmark{a,b} \\
      \omega & \text{deg} & 89.6\pm1.2~\tablefootmark{c} \\
      \hline
      \noalign{\smallskip}
      \multicolumn{4}{c}{\text{MCMC Transit Fitting Results}} \\
      \hline
      \text{Parameter} & \text{V1} & \text{V2} & \text{V3} \\
      \hline
      t_0 & 0.036^{+0.004}_{-0.004} & 0.039^{+0.005}_{-0.005} & 0.014^{+0.011}_{-0.015} \\
      R_\mathrm{p}/R_\mathrm{s} & 0.104^{+0.005}_{-0.005} & 0.107^{+0.007}_{-0.007} & 0.103^{+0.007}_{-0.006} \\
      (u_1, u_2) & (0.335, 0.392) & (0.381, 0.395) & (0.394, 0.427) \\
      \hline
   \end{array}
   $$
   \tablefoot{
   Symbol definitions: $P$: Orbital period. $a/R_\mathrm{s}$: Semi-major axis in stellar radii. $i$: Orbital inclination. $e$: Orbital eccentricity. $\omega$: Orbital obliquity. $t_0$: Time of inferior conjunction (days). $R_\mathrm{p}/R_\mathrm{s}$: Planet radius in stellar radii. $(u_1, u_2)$: Quadratic limb darkening coefficients. \\
   \tablefoottext{a}{\citet{Lendl2020}} 
   \tablefoottext{b}{\citet{Anderson2018}}
   \tablefoottext{c}{\citet{Deline2022}}
   }
\end{table}

\subsection{Mass-loss and stellar wind constraints}

The stellar wind environment at WASP-189b is estimated using escape velocity ($v_{\mathrm{esc}} $$=$$ \sqrt{2GM_\mathrm{s}/R_\mathrm{s}}$), which yields $v_{\mathrm{esc}}$$\approx$$573$ km~s$^{-1}$ for WASP-189. Based on line-driven wind theory \citep{CAK1975}, the terminal wind speed is typically $v_{\mathrm{sw}} = [1, 3]~v_{\mathrm{esc}}$, corresponding to 573--1719~km~s$^{-1}$. In the intermediate- and high-speed regimes of this range, $v_{\mathrm{sw}}$ exceeds the solar wind upper limit (850~km~s$^{-1}$; \citealt{Alissandrakis2021,Pevtsov2021}). Theoretical models further suggest that fast-rotating A-type stars, like WASP-189 ($P_\mathrm{rot}$$\approx$$1.24$ days; \citealt{Lendl2020,Deline2022}), can drive even higher wind speeds, consistent with the adopted simulation values \citep{Johnstone2015}.

A-type stars' coronal activity, relevant for wind properties, is not always well constrained. WASP-189's $T_\mathrm{eff}$$=$$7996$~K is near the threshold for weak coronal emission \citep{Gunther2022}, analogous to Altair (A7 V) with $T_{\mathrm{cor}}$$\sim$$10^6$ K \citep{Gudel2004}. This coronal temperature is used in our subsequent density calculations. The number density at the planetary orbital distance was then estimated following \citet{Vidotto2010}:

\begin{equation}
\frac{n_{\mathrm{obs}}}{n_0}
= \exp \left[
  \frac{GM_\mathrm{s}/R_\mathrm{s}}{k_\mathrm{B} T_\mathrm{cor}/m}
    \left( \frac{R_\mathrm{s}}{R_{\mathrm{orb}}} - 1 \right)
  + \frac{2 \pi^2 R_\mathrm{s}^2 / P_\mathrm{s}^2}{k_\mathrm{B} T_\mathrm{cor}/m}
    \left( \frac{R_{\mathrm{orb}}^2}{R_\mathrm{s}^2} - 1 \right)
\right]~.
\end{equation}Here, $n_0$ is the coronal base number density, derived from mass continuity
\(\dot{M} = 4\pi R_\mathrm{s}^2 \rho_0 v_0\) \citep{Lanz1992} with
\(\dot{M} \sim 2 \times 10^{-10}\,M_\odot\,\mathrm{yr}^{-1}\).
The base wind speed is taken as the isothermal sound speed,
\(v_0 \approx \sqrt{k_\mathrm{B} T_\mathrm{b} / (\mu m_\mathrm{p})}\)
\citep{Lamers1999}, with \(T_\mathrm{b} \sim 8000\,\mathrm{K}\) and
\(\mu \sim 0.6\).
Then, \(\rho_0 = \dot{M} / (4\pi R_\mathrm{s}^2 v_0)\) and
\(n_0 = \rho_0 / (\mu m_\mathrm{p})\).

Applying these relations, we estimate the stellar wind density at the orbital distance of WASP-189\,b to be $\rho_{\rm sw} \approx 5.32 \times 10^{-13}~\mathrm{kg\,m^{-3}}$. The condition for bow-shock formation is set by the fast-mode magnetosonic Mach number, $M_\mathrm{F} = v_{\mathrm{sw}} / C_{\mathrm{f}}$ \citep{Lai2024}, where the fast magnetosonic speed is $C_{\mathrm{f}} = \sqrt{C_\mathrm{A}^2 + C_\mathrm{s}^2}$. Here $C_\mathrm{A} = B / \sqrt{\mu_0 \rho}$ is the Alfvén speed, with $\mu_0$ denoting the vacuum permeability, and $C_\mathrm{s} = \sqrt{\gamma k_\mathrm{B} T_\mathrm{cor} / (\mu m_\mathrm{p})}$ is the adiabatic sound speed for a monatomic, fully ionised gas, where $\gamma = 5/3$, $T_\mathrm{cor}$ is the coronal temperature, $\mu$ is the mean molecular weight, and $m_\mathrm{p}$ is the proton mass.

To achieve \(M_{\mathrm{F}}>1\), a plasma density \(\gtrsim 10^{-11}\,\mathrm{kg\,m^{-3}}\) is required. Since the ambient wind is less dense, local enhancement from the planetary upper atmosphere and magnetosheath compression is plausible. For UHJs, the wind–atmosphere interface is diffuse and variable, so \(\rho \sim \rho_{\mathrm{atm}}\) is assumed for the stellar-wind downstream region here, estimated via \(\rho=\mu\,m_{\mathrm p}\,p/(k_{\mathrm B}\,T_{\mathrm p})\) \citep{Rapp-Kindner2024} with \(T_{\mathrm p}=3435~\mathrm{K}\) and \(p=1~\mathrm{mPa}\), giving \(\rho_{\mathrm{atm}}\sim 4.59\times 10^{-11}\,\mathrm{kg\,m^{-3}}\)—above the shock-formation threshold and consistent with NUV variability under suitable wind conditions.

\section{Magnetohydrodynamic simulations}

To investigate the connection between WASP-189b’s early NUV ingress and the formation of a MBS, we mapped the fast-mode Mach number ($M_\mathrm{F}$) as a function of plasma density and stellar wind velocity using 2D ideal magnetohydrodynamic (MHD) simulations (Appendix~\ref{append_2D_sim}). The simulations adopt wind speeds of $1$–$3~v_\mathrm{esc}$, plasma densities of either $\rho_\mathrm{atm}$ or $\rho_\mathrm{sw}$, and stellar and planetary magnetic fields of 79.05~G and 60.22~G, respectively (Appendices~\ref{HJ_B_surf} and \ref{app:k2012}).

\begin{figure}[!htp]
\centering
\includegraphics[width=0.5\textwidth]{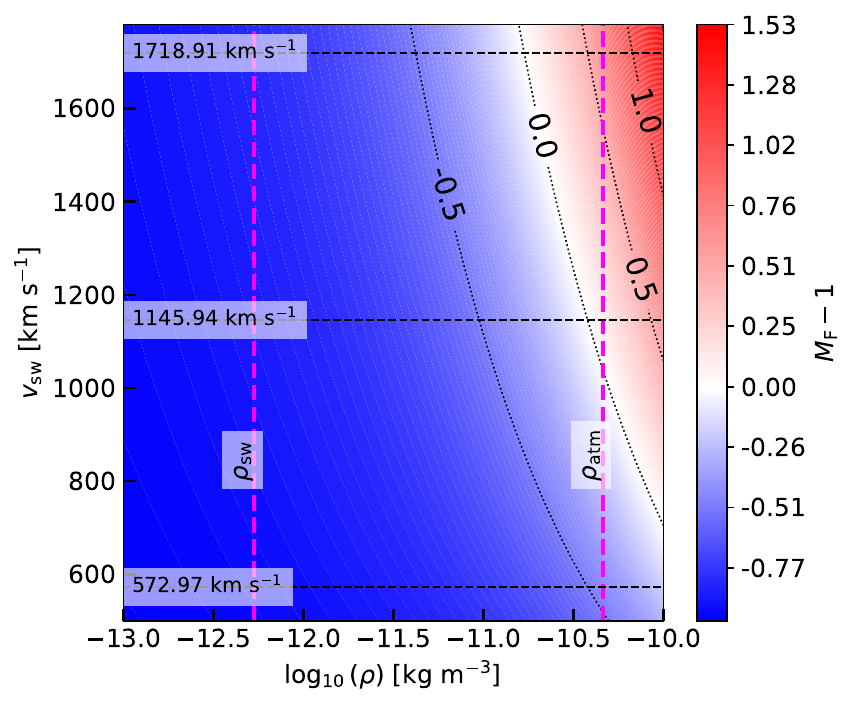}
\caption{Fast-mode magnetosonic Mach number under WASP-189b-like stellar wind conditions.
The colour map shows $M_\mathrm{F} - 1$ as a function of $\log_{10}(\rho)$ and $v_{\mathrm{sw}}$, for a stellar magnetic field of 79.05~G. White contours mark the $M_\mathrm{F} = 1$ boundary. Vertical magenta lines indicate estimated planetary upper atmosphere and stellar wind densities; horizontal dashed lines denote wind speeds of $1$–$3v_\mathrm{esc}$. Regions where $M_\mathrm{F} > 1$ are conducive to bow shock formation.}\label{fig2}
\end{figure}

\begin{figure}[!htp]
\centering
\includegraphics[width=0.5\textwidth]{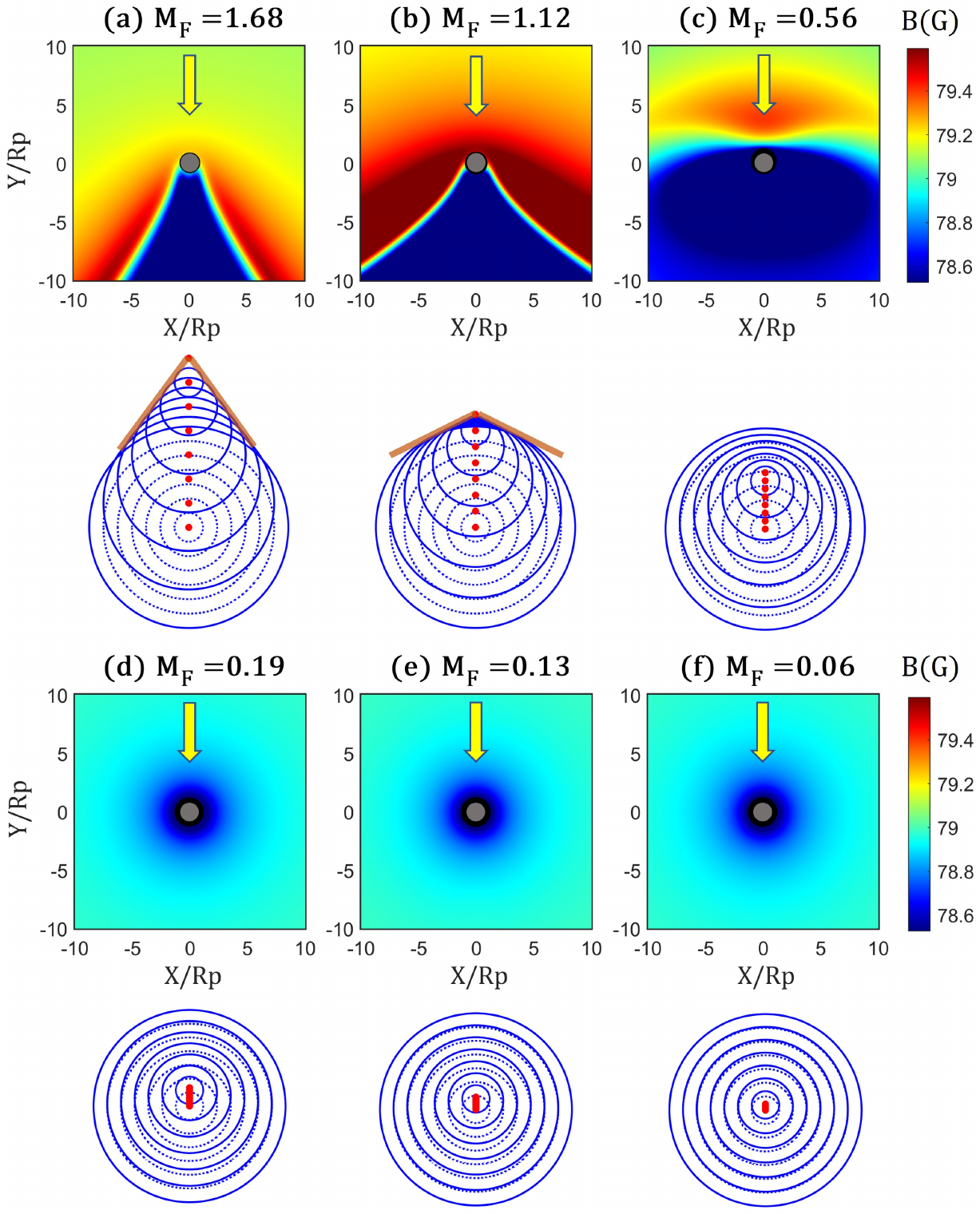}
\caption{2D MHD simulations of bow shock formation for WASP-189b under varying stellar wind conditions.
Six combinations of stellar wind speed ($v_\mathrm{sw}=1$–$3v_\mathrm{esc}$) and plasma density ($\rho_\mathrm{atm}$ or $\rho_\mathrm{sw}$), with a stellar magnetic field of 79.05~G, are shown. Top: Magnetic field strength (contours) and wind streamlines (yellow arrows); black circle marks the magnetopause stand-off distance ($R_\mathrm{mp}=1.09$–$1.58~R_\mathrm{p}$). Bottom: Fast-mode wavefronts and shock cones (brown), showing bow shock formation only for $M_\mathrm{F}>1$. Panels (a) and (b) demonstrate that high $v_\mathrm{sw}$ combined with high $\rho$ is required for a distinct bow shock.}\label{fig3}
\end{figure}

Figure~\ref{fig2} shows that $M_\mathrm{F}$ exceeds unity—allowing bow shock formation—only when both stellar wind velocity and plasma density are sufficiently high, specifically for $v_\mathrm{sw} \gtrsim 2v_\mathrm{esc}$ and $\rho = \rho_\mathrm{atm}$. For typical stellar wind densities ($\rho_\mathrm{sw}$), the flow remains sub-magnetosonic and no shock is produced, consistent with MHD theory and prior results~\citep{Alexander2015}. Figure~\ref{fig3} further illustrates that clear, well-defined bow shocks emerge only under these high-$v_\mathrm{sw}$, high-$\rho$ conditions; lower wind speeds generate only broad, diffuse density enhancements. The maximum spatial extent of these structures (up to $\sim$$5~R_\mathrm{p}$) is comparable to the $\sim$$3.7~R_\mathrm{p}$ absorption region inferred from the EI, linking the $M_\mathrm{F}$ parameter space (Fig.~\ref{fig2}) with the simulated density morphologies (Fig.~\ref{fig3}).

Overall, while our simulations confirm that MBS formation is dynamically feasible for WASP-189b, the detectability of such structures in the NUV is constrained by shock cooling physics. Detectable NUV absorption requires that the radiative cooling time ($\tau_{\rm cool}$) be shorter than the shock crossing time ($\tau_{\rm cross}$). As shown in Appendix~\ref{shock_time}, this criterion is satisfied only for moderate stellar wind speeds ($v_{\rm sw} = 572.97~\mathrm{km\,s^{-1}}$) when the shock thickness lies within $2.45$–$2.94~R_{\rm p}$ (Appendix~\ref{shock_thickness}), enabling rapid cooling and sufficient NUV opacity. 

At higher wind speeds, a bow shock can still form due to supersonic flow; however, the post-shock region remains too hot for efficient NUV absorption, rendering it effectively invisible at these wavelengths. If the stellar wind subsequently transitions from a high-speed to a low-speed regime, $\tau_{\rm cool}$ decreases while the compressed plasma in the pileup region remains dense. Under these conditions, the cooled, optically thick sheath ahead of the magnetosphere can persist long enough to be observed in the NUV, potentially producing the EI signature (Table~\ref{tab:cooling_times}). This scenario links the dynamic MBS formation seen in our simulations with the radiative conditions necessary for observational detection.

\section{Discussions and conclusions}

CUTE NUV transit observations of WASP-189b reveal a possible EI in V3 that was absent in V1 and~V2.\ This suggests episodic, potentially wind-driven variability in the planet’s upper atmosphere. Similar anomalies in other hot Jupiters have been attributed to magnetospheric interactions or bow shocks \citep{Vidotto2010,Llama2011,Sreejith2023}, although our dataset does not confirm their persistence.

We assessed whether the observed signature could arise from MBS effects. Our MHD simulations show that classical MBS formation ($M_\mathrm{F} > 1$) requires both a high stellar wind velocity and sufficient plasma density. However, shock-cooling constraints indicate that only moderate winds ($v_{\rm sw} \approx v_\mathrm{esc}$) enable rapid post-shock cooling and adequate NUV opacity. In this regime, a transient high-density pileup can form within $\lesssim 5~R_\mathrm{p}$ ahead of the planet, consistent with the $\sim 3.65~R_\mathrm{p}$ absorption extent inferred from V3.

Moreover, comparison of the shock cooling time and shock crossing time indicates that NUV detectability is suppressed at the moment of MBS formation (Table~\ref{tab:cooling_times}): the post-shock plasma remains too hot for efficient absorption, and by the time it cools sufficiently, the shock structure has already evolved from its initial state. If this evolution is followed by a transition from high to low stellar-wind speeds, EI visibility can increase substantially because the shorter $\tau_{\rm cool}$ at lower $v_{\mathrm{sw}}$ allows the residual compressed plasma to cool while maintaining a high density. The pressure budget at $R_{\mathrm{mp}}$ (Table~\ref{tab:k2012_combined}) supports this scenario: in the low-density case, magnetic pressure dominates over ram pressure across all wind speeds, and in the high-density case it dominates at low wind speeds. Such magnetic pileup ahead of the magnetosphere can facilitate dayside reconnection, channelling cooled plasma into the compressed sheath and increasing its NUV opacity—thereby favouring EI detection following a fast-to-slow wind transition.

In summary, while the true origin of the CUTE V3 offset remains uncertain, our study examines the possibility that NUV EI signatures in WASP-189b could arise from stellar-wind compression of the UHJ's magnetosphere and associated plasma pileup. Our Bayesian analysis yields moderate evidence consistent with an EI feature in V3. This scenario highlights the value of coordinated, high-cadence UV and X-ray monitoring to better constrain stellar-wind conditions, assess planetary magnetic field strengths, and evaluate the predicted sensitivity of EI visibility to variations in wind properties.

\begin{acknowledgements}

This work was supported by the Upper Air Dynamics Laboratory and CAPE at National Central University (NCU), with funding from NSTC grants 114-2917-I-564-044, 113-2111-M-008-007, 112-2811-M-008-072, 113-2811-M-008-001, 114-2111-M-008-008, and the Higher Education SPROUT grant from Taiwan’s Ministry of Education. H.J.H. acknowledges eSSENCE (eSSENCE@LU 9:3), the Swedish National Research Council (2023-05307), the Crafoord Foundation, and the Royal Physiographic Society of Lund. A.J. is supported by the Carlsberg Foundation (FIRSTATMO). We thank the CUTE team at the University of Colorado and the NCU Upper Air Dynamics Laboratory for data provision and downlink support.

Y.D. thanks A. Johansen (U. Copenhagen) for hosting, G. Chen (PMO, CAS) for feedback, C.-H. Lin, J.-Y. Liu, Y.-C. Wen (NCU) for planetary and magnetospheric discussions, U.G. Jørgensen (NBI) for spectral advice, Y. Tian (Sejong U.) for UHJ discussions, and Y.-C. Chiu and R.-T. Chen for CUTE ground station support. An anonymous COSPAR 45 attendee is thanked for suggesting this investigation.

Data availability:
CUTE NUV transit data are available from the mission PI (K. France) on request. CHEOPS optical data are public via the ESA CHEOPS archive (\url{https://www.ssdc.asi.it/cheops/}) and as supplementary files in \citet{Lendl2020}. Processed data and analysis scripts are available from the corresponding author upon request.

\end{acknowledgements}

\bibliographystyle{aa}
\bibliography{sn-bibliography}

\begin{appendix}

\section{Phase-dependent NUV spectral variability around Mg~\textsc{ii}}

Spectral analysis around Mg~\textsc{ii} (2765.6–2834.6~\AA) reveals time-dependent anomalies in V3 (e.g. emission at ingress, strong absorption post-egress) and V2, but not in V1. The rapid variability suggests temporal changes in the planet’s upper atmosphere or magnetospheric environment, though precise identification is limited by the S/N.

\begin{figure}[h!]
\centering
\includegraphics[width=0.5\textwidth]{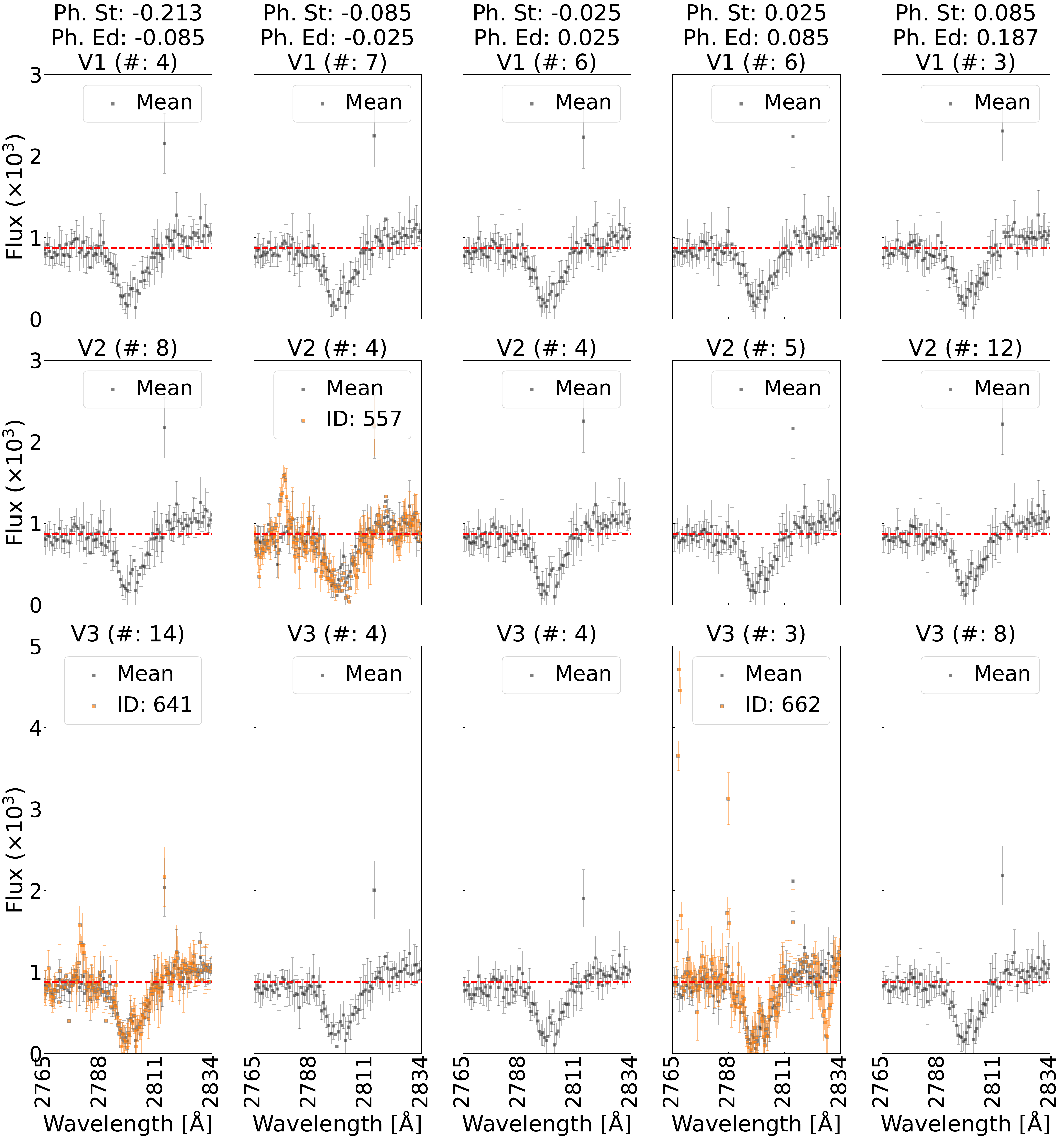}
\caption{Phase-resolved spectral analysis of WASP-189 b around the Mg II region (2765.6-2834.6~\AA). Each panel shows the flux variation (counts, not normalised) as a function of wavelength for different observational visits (V1, V2, and V3) and phase bins (Ph). The symbol \# indicates the number of samples in each bin, while St and Ed denote the start and end phases, respectively. The grey markers represent the mean flux profile across multiple observations, with error bars indicating the standard deviation. Three specific events (557, 641, and 662) that deviate from the mean trend  are highlighted in orange. The dashed red line denotes the median continuum level for each visit. These results provide insights into the temporal and phase-dependent variations in the observed spectral features, offering constraints on the atmospheric and astrophysical processes affecting WASP-189 b.}\label{append_mgii}
\end{figure}

\section{Early ingress and the Bayes factor}\label{EI_Bayes}

\begin{figure}[h!]
\centering
\includegraphics[width=0.49\textwidth]{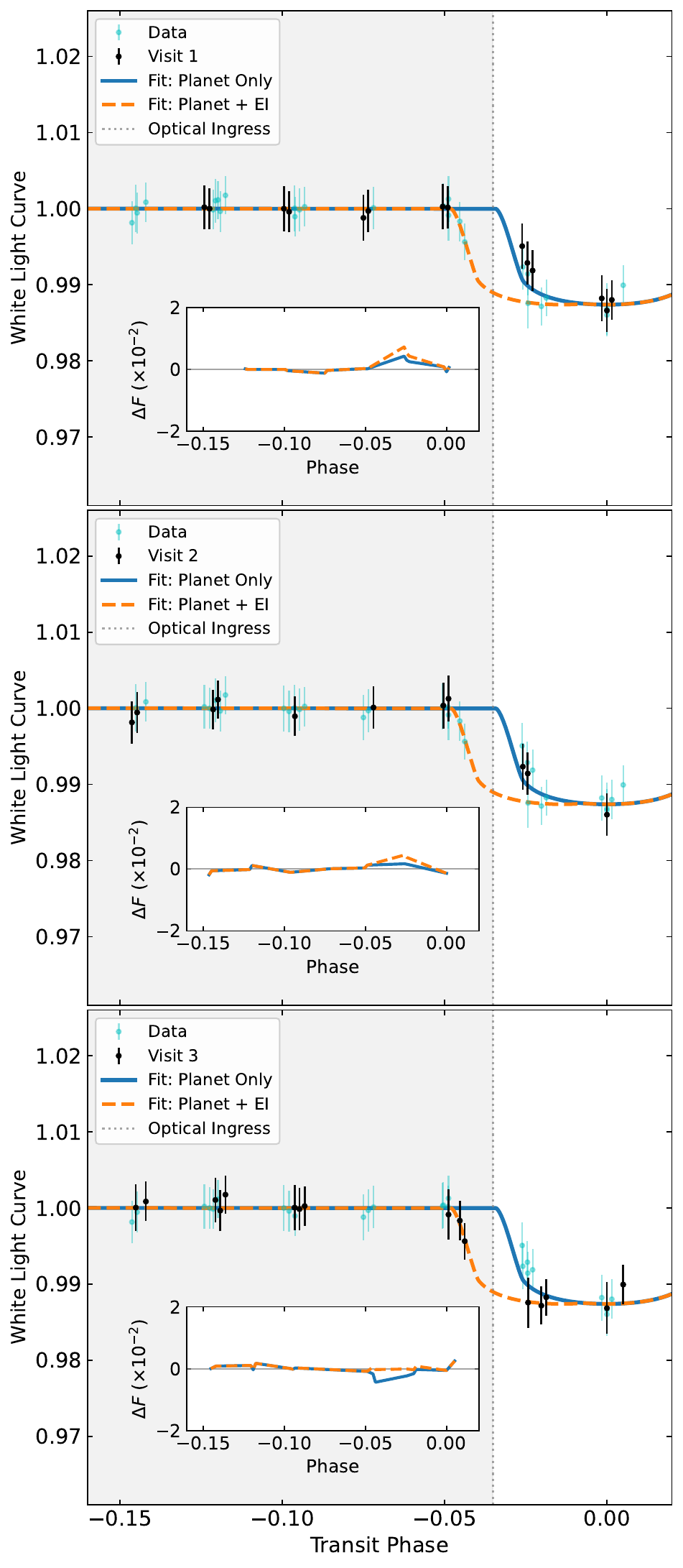}
\caption{Per-visit comparison between the nominal transit model and the phase-shifted EI model. Each row corresponds to one WASP-189b transit (V1-V3). The main panels display the observed white light curves (cyan for all visits, black for the highlighted visit) together with the best-fitting models: the nominal planet-only transit (solid blue) and the planet+EI model (dashed orange). The insets show the corresponding residuals, $\Delta F$, relative to each model. The EI model is constructed as the pointwise minimum of the base transit and a phase-shifted replica, providing a diagnostic test for the presence of an earlier ingress.}
\label{append_EI_bayes}
\end{figure}

To test for the presence of an EI absorption feature in the NUV light curves, we constructed a phase-shifted replica of the fiducial transit model using the \texttt{batman} package \citep{Kreidberg2015}. As illustrated in Fig.~\ref{append_EI_bayes}, the observed light curves for each visit (grey for all visits, black for the highlighted one) are compared against two models: the nominal planet-only transit (blue solid line) and the planet+EI model (dashed orange line). The replica model adopts the same geometric parameters as the optical transit, and the only additional degree of freedom is a phase offset $\Delta \phi$, which shifts the centre of the replica earlier relative to the nominal transit (negative $\Delta \phi$ corresponds to earlier ingress). The composite flux is then defined pointwise as
\begin{equation}
    F_{\mathrm{model}}(\phi) = \min \left[ F_{\mathrm{base}}(\phi), \,
    F_{\mathrm{shift}}(\phi + \Delta \phi) \right],
\end{equation}
where $F_{\mathrm{model}}(\phi)$ is the combined flux at orbital phase $\phi$, 
$F_{\mathrm{base}}(\phi)$ is the nominal planet-only transit, 
$F_{\mathrm{shift}}(\phi + \Delta \phi)$ is the shifted replica, 
$\phi$ denotes the orbital phase relative to mid-transit ($\phi=0$), and 
$\Delta \phi$ is the free phase-lead parameter.

Bayesian model comparison was carried out using the Bayes factor, defined as the ratio of the marginal likelihoods (Bayesian evidence) of two competing models \citep{Jeffreys1948, Kass1995, Heller2024}. The evidence, $\mathcal{Z,}$ was estimated with nested sampling, yielding
\begin{equation}
    2\ln B = 2 \left( \log \mathcal{Z}_{\mathrm{EI}} - \log \mathcal{Z}_{\mathrm{base}} \right),
\end{equation}
where $\mathcal{Z}_{\mathrm{EI}}$ and $\mathcal{Z}_{\mathrm{base}}$ denote the evidence for the planet+EI and planet-only models, respectively. Positive values of $2\ln B$ favour the EI hypothesis, and the results can be interpreted according to the Jeffreys scale. 
In practice, $\log \mathcal{Z}_{\mathrm{EI}}$ was obtained from nested sampling (via \texttt{UltraNest}). For the baseline planet-only model, we used its maximum Gaussian log-likelihood as a proxy for $\log \mathcal{Z}_{\mathrm{base}}$.

Applying this framework to the three NUV visits, and restricting the analysis to the transit phase interval $-0.05 \leq \phi \leq -0.01$ in order to isolate the ingress region, we find negative Bayes factor values for the first two ($2\ln B \approx -7.87$ and $-4.46$), indicating no preference for the EI model. By contrast, the third visit yields $2\ln B \approx 3.78$, corresponding to moderate statistical support for an EI feature within this pre-ingress window.

\section{Hot Jupiter surface magnetic field}\label{HJ_B_surf}

We estimated the surface magnetic field strength $B_\mathrm{surf}$ of WASP-189b using the scaling approach described by \citet{Dietrich2022}, which is based on the Christensen dynamo scaling law \citep{Christensen2009}. This method relates a planet’s magnetic field to its internal heat flux under the assumption that the magnetic energy is set by the convective power available in the dynamo region. Their model assumes that the dynamo operates in a rapidly rotating, electrically conducting shell, that convection is driven by the internal heat flux, $q_\mathrm{int}$, and that the ohmic dissipation is small compared to the available convective power. The theory builds on magnetostrophic balance and energy-based scaling arguments \citep{Christensen2009}, which have been successfully applied to both Solar System and exoplanetary giant planets.

For WASP-189b, we estimated $q_\mathrm{int}$ from its equilibrium temperature ($T_\mathrm{eq}$) using the empirical relation \citep{Thorngren2018, Sarkis2021} as adopted by \citet{Dietrich2022}:
\begin{equation}
    T_\mathrm{int} = 0.39\, T_\mathrm{eq} \, \exp\left[ -\frac{\left( \log_{10}(\sigma_\mathrm{SB} T_\mathrm{eq}^4) - 6.14 \right)^2}{1.095} \right],
\end{equation}
\begin{equation}
    q_\mathrm{int} = \sigma_\mathrm{SB} T_\mathrm{int}^4,
\end{equation}
where $T_\mathrm{int}$ is the internal temperature, $T_\mathrm{eq}$ is the planetary equilibrium temperature, and $\sigma_\mathrm{SB}$ is the Stefan–Boltzmann constant. We adopted Jupiter’s internal heat flux $q_{\mathrm{int,J}}=5.4\,\mathrm{W\,m^{-2}}$ and surface magnetic field $B_\mathrm{J}=4.17\times 10^{-4}\,\mathrm{T}$ as reference values~\citep{Dietrich2022}.

The surface magnetic field was then obtained from the scaling law:
\begin{equation}
    B_\mathrm{surf} = B_\mathrm{J} \left( \frac{R_\mathrm{J}}{R_\mathrm{p}} \right)^{1/2}
                           \left( \frac{M_\mathrm{p}}{M_\mathrm{J}} \right)^{1/6}
                           \left( \frac{q_\mathrm{int}}{q_{\mathrm{int,J}}} \right)^{1/3},
\end{equation}
where $M_\mathrm{J}$ and $R_\mathrm{J}$ are Jupiter’s mass and radius. This approach avoids assumptions on planetary age, instead inferring $q_\mathrm{int}$ from $T_\mathrm{eq}$, which is particularly suitable for highly irradiated planets such as WASP-189b. Applying this method to WASP-189b yields an estimated surface magnetic field strength of $B_\mathrm{surf} \approx 60.22\,\mathrm{G}$.

\section{Magnetopause and pressure budget}\label{app:k2012}

Direct measurements of WASP-189b’s magnetic field are unavailable currently; however, rapidly rotating A-type stars (such as WASP-189, $P_\mathrm{rot}$=$1.24$~days; \citealt{Lendl2020, Deline2022}) are expected to possess strong surface fields, often exceeding $7500~\mathrm{G}$ \citep{Mathys2017}. Assuming a dipole geometry, the stellar field at the planet’s orbit ($r$=$ 0.05$~AU) follows $B(r)$=$ B_0 (R_0/r)^3$, giving $\sim$$79.05$~G for our MHD models. 

The dayside magnetopause stand-off distance of WASP-189b is computed using the magnetodisc–dipole formulation of \citet{Khodachenko2012}, which is based on a paraboloid magnetospheric model and accounts for the extended magnetodiscs of hot Jupiters that may arise from the outflow of ionised particles in a hydrodynamically expanding upper atmosphere. In this framework, the inner edge of the magnetodisc (the Alfvénic radius) is given by
\begin{equation}
    R_{\rm A} = R_{\rm p} \left( \frac{2\pi\,\delta\theta\,B_{\rm d0}^2\,R_{\rm p}}{\mu_0\,\omega_{\rm p}\,\dot{M}} \right)^{1/5},
\end{equation}
where \(B_{d0}\) denotes the planetary equatorial surface field (we adopted \(60.22\,\mathrm{G}\)), and \(\delta\theta\) is the angular half-thickness of the magnetodisc; we used $\delta\theta \approx 0.1\,\mathrm{rad}$ as a representative thin-disc value, motivated by Jovian magnetodisc studies that infer angular thicknesses of order a few degrees to $\sim10^{\circ}$ from field and particle data (e.g. \citealp{Khurana2005,Connerney2020}). Since WASP-189b is tidally locked, the spin period is equivalent to the orbital period, \(2.724~{\rm days} \;(\approx 2.35\times 10^{5}~{\rm s})\), giving the planetary angular velocity
\begin{equation}
    \omega_{\rm p} = \frac{2\pi}{P_{\rm orb}} \approx 2.67\times 10^{-5}~{\rm rad\,s^{-1}}.
\end{equation}

The substellar pressure balance at \(r = R_{\mathrm{mp}}\) is
\begin{equation}
    \kappa^2 \frac{\big[B_{\rm d}(r) + B_{\rm MD}(r)\big]^2}{2\mu_0} + p_{\rm MD}(r) = p_{\rm sw},
\end{equation}
where the dipole component is \(B_{\rm d}(r) = B_{\rm d0} (R_{\rm p}/r)^3\), the magnetodisc component is \(B_{\rm MD}(r) = B_{\rm d0} R_{\rm p}^3 / (R_{\rm A} r^2)\), and the magnetodisc plasma pressure is
\begin{equation}
    p_{\rm MD}(r) = \frac{B_{\rm d0}^2 R_{\rm p}^6}{2\mu_0\,R_{\rm A}^2\,r^4}.
\end{equation}
The parameter \(\kappa = 2f_0\) represents the current-closure geometry factor, with \(f_0 \simeq 1.22\) adopted here~\citep{Khodachenko2012}, giving \(\kappa \simeq 2.44\)~\citep{Alexeev1997, Alexeev2003}.  
The stellar-wind total pressure is
\begin{equation}
    p_{\rm sw} = \rho_{\rm sw} v_{\rm sw}^2 + n_{\rm sw} k_{\rm B} T_{\rm sw} + \frac{B_{\rm sw}^2}{2\mu_0},
\end{equation}
where \(n_{\rm sw} = \rho_{\rm sw} / m_{\rm p}\) is the number density, \(T_{\rm sw}\) is the stellar-wind temperature, and \(B_{\rm sw}\) is the stellar-wind magnetic field.

The planetary thermal outflow was treated as an isothermal wind with temperature \(T_{\rm th} = 1.5\times 10^4\)~K~\citep{Sreejith2023} and sound speed \(c_s = \sqrt{k_{\rm B} T_{\rm th} / m_{\rm p}}\); it was used to evaluate \(R_{\rm A}\) and the planetary-side thermal pressure:
\begin{equation}
    p_{\rm th}(r) = \frac{\dot{M} c_{\rm s}}{4\pi r^2}.
\end{equation}
The planetary surface equatorial magnetic field is set to \(B_{\rm d0} = 60.22~{\rm G}\) (Appendix~\ref{HJ_B_surf}). For the stellar wind we used  $10^6$~K for the $T_{\rm sw}$ such that $T_{\rm sw} = T_{\rm cor}$.

Table~\ref{tab:k2012_combined} summarises the stand-off distance \(R_{\mathrm{mp}}\) from the \citet{Khodachenko2012} solution, and the pressure components for two plasma densities. The planetary and stellar magnetic pressures are estimated to be $p_{\rm B}\sim14.43$ Pa and $p_{B,\rm sw}\sim24.86$ Pa, respectively. The listed pressures  include the planetary and stellar thermal pressures at the magnetopause distance ($p_{\rm th}$ and $p_{\rm th,\rm sw}$), as well as the stellar ram pressure ($p_{\rm ram} = \rho_{\rm sw} v_{\rm sw}^2$). Here $p_{\rm ram}$ represents the dynamic pressure exerted by the bulk motion of the stellar wind plasma, which acts to compress the planetary magnetosphere in competition with magnetic and thermal pressures.

\begin{table}[h!]
\centering
\caption{Magnetopause stand-off and pressure budget for two plasma densities.}\label{tab:k2012_combined}
\begin{tabular}{lccc}
\hline\hline
$v_{\mathrm{sw}}$ (km\,s$^{-1}$) & 1718.91 & 1145.94 & 572.97 \\
\hline
\multicolumn{4}{c}{\textbf{$\rho = 4.59\times 10^{-11}~{\rm kg~m^{-3}}$}} \\
\hline
$R_{\mathrm{mp}}~(R_{\rm p})$        & 1.09 & 1.24 & 1.44 \\
$p_{\rm th}$ (Pa)  & 0.70 & 0.54 & 0.40 \\
$p_{\rm ram}$ (Pa)           & 135.62 & 60.27 & 15.07 \\
$p_{\rm th,sw}$ (Pa)           & 0.38 & 0.38 & 0.38 \\
\hline
\multicolumn{4}{c}{\textbf{$\rho = 5.32\times 10^{-13}~{\rm kg~m^{-3}}$}} \\
\hline
$R_{\mathrm{mp}}~(R_{\rm p})$     & 1.57 & 1.58 & 1.58 \\
$p_{\rm th}$ (Pa)  & 0.34 & 0.33 & 0.33 \\
$p_{\rm ram}$ (Pa)           & 1.57 & 0.70 & 0.17 \\
$p_{\rm th,sw}$ (Pa)           & $4.39\times10^{-3}$ & $4.39\times10^{-3}$ & $4.39\times10^{-3}$ \\
\hline
\end{tabular}
\end{table}

While $R_{\mathrm{mp}}$ sets the size of the planetary obstacle in our MHD models, the pressure components in Table~\ref{tab:k2012_combined} are evaluated at the magnetopause location. For the low stellar-wind density case ($\rho = 5.32\times 10^{-13}~{\rm kg~m^{-3}}$), $p_{\rm ram} < p_{\rm B,{\rm sw}}$ for all three wind speeds, indicating strong magnetic confinement regardless of flow velocity. In contrast, for the high-density case, which in our assumptions corresponds to conditions in the stellar-wind downstream region ($\rho = 4.59\times 10^{-11}~{\rm kg~m^{-3}}$), $p_{\rm ram} > p_{\rm B,{\rm sw}}$ at moderate and high wind speeds, but $p_{B,{\rm sw}}$ dominates at low wind speed.  

Physically, when $p_{\rm ram} > p_{\rm B,{\rm sw}}$, the magnetopause is compressed primarily by dynamic pressure from the wind, potentially narrowing open-field regions and reducing polar outflow efficiency. When $p_{\rm B,{\rm sw}}$ dominates, the stellar magnetic field governs the interaction, promoting stronger magnetic pileup ahead of the magnetosphere. Such pileup can enhance magnetic reconnection rates, facilitating the conversion of magnetic energy into heat and potentially increasing the column density in the compressed sheath—conditions that may favour detectability of NUV EI signatures.

\section{Details of the 2D MHD simulation}\label{append_2D_sim}

Two-dimensional ideal MHD simulations were carried out using an in-house Fortran code with a second-order Lax–Wendroff scheme \citep{Richtmyer1967} for time integration and an alternating direction implicit method \citep{Press1988} for stable numerical diffusion. The simulations solve the standard set of MHD equations in Cartesian coordinates: mass continuity, momentum conservation, energy conservation, and the induction equation. Explicitly, these are
\begin{align}
&\frac{\partial \rho}{\partial t} + \nabla \cdot (\rho \mathbf{v}) = 0, \\
&\frac{\partial (\rho \mathbf{v})}{\partial t} + \nabla \cdot \left[\rho \mathbf{v}\mathbf{v} + \left(p + \frac{B^2}{2\mu_0}\right)\mathbf{I} - \frac{\mathbf{B}\mathbf{B}}{\mu_0}\right] = 0, \\
&\frac{\partial E}{\partial t} + \nabla \cdot \left[\left(E + p + \frac{B^2}{2\mu_0}\right)\mathbf{v} - \frac{\mathbf{B}(\mathbf{B}\cdot\mathbf{v})}{\mu_0}\right] = 0, \\
&\frac{\partial \mathbf{B}}{\partial t} + \nabla \times (\mathbf{v} \times \mathbf{B}) = 0,
\end{align}
\noindent
where $\rho$ is the mass density, $\mathbf{v}$ the plasma velocity, $\mathbf{B}$ the magnetic field, $p$ the thermal pressure, and $\mu_0$ the permeability of free space. The total energy density is $E = \frac{p}{\gamma - 1} + \frac{1}{2} \rho v^2 + \frac{B^2}{2\mu_0}$, with $\gamma$ the adiabatic index.

All variables are assumed uniform in the $z$-direction, such that $\partial / \partial z = 0$. The computational domain spans $20~R_p \times 20~R_p$, with a uniform grid spacing of $0.1~R_p$. Uniform boundary conditions are imposed at all edges. The electric field is eliminated using the ideal MHD Ohm’s law, and the system is advanced in time using the above Lax–Wendroff alternating direction implicit algorithm. This scheme allows us to self-consistently model the interaction of the stellar wind and planetary magnetosphere, capturing the formation of shocks, pileup regions, and the global plasma environment relevant to exoplanet transit phenomena.

\section{Shock thickness}\label{shock_thickness}

We estimated the thickness of the shocked sheath, $\sigma_\mathrm{sh}$, from the observed EI timing offset. The projected orbital velocity during transit was derived from the transit geometry described in \citet{Damiani2011} and can be written as
\begin{equation}
    v_\perp = \frac{2 R_s \sqrt{1 - b^2}}{T_\mathrm{dur}},
\end{equation}
where $R_s$ is the stellar radius, $b$ is the impact parameter ($b=0.4537$; \citealt{Lendl2020}), and $T_\mathrm{dur}$ is the total transit duration. The projected stand-off distance between the absorbing structure and the planetary disc is then
\begin{equation}
    \Delta s = v_\perp \, \Delta t ,
\end{equation}
where $\Delta t$ is the EI time offset measured from our light curves ($\sim$31.54~min). Assuming the first detectable absorption occurs at the bow shock nose, the shock thickness is
\begin{equation}
    \sigma_\mathrm{sh} = \Delta s - (R_\mathrm{mp} - R_\mathrm{p}).
\end{equation}
This formulation accounts for the distance from the planetary surface to the magnetopause before subtracting from the observed leading distance. For the three stellar wind cases considered here, the resulting $\sigma_\mathrm{sh}$ values range from 2.45 to 2.94~$R_\mathrm{p}$.

\section{Shock cooling and crossing time}\label{shock_time}

The post-shock temperature is estimated as $T_{\rm ps} \sim 2\times10^5 (v_{\rm sh}/100~\mathrm{km\,s^{-1}})^2$~K \citep{delValle2022}, where the shock velocity is $v_{\rm sh} \approx \sqrt{v_{\rm sw}^2 + v_{\rm orb}^2}$ and the planetary orbital velocity is $v_{\rm orb} \approx 188.86~\mathrm{km\,s^{-1}}$. The post-shock density is taken as $n_{\rm ps} = 4\rho_{\rm atm} / (\mu m_{\rm p})$ under the assumption of a fourfold compression \citep{Lai2019}. Cooling times are then calculated via
\begin{equation}
    \tau_{\rm cool} = \frac{3k_\mathrm{B} T_{\rm ps}/2}{n_{\rm ps} \Lambda(T_{\rm ps})},
\end{equation}
where $\Lambda(T_\mathrm{ps}) \approx \Lambda_\odot(T_\mathrm{ps})\,10^{[\mathrm{Fe}/\mathrm{H}]}$ is the metallicity-scaled radiative cooling function, adopting $[\mathrm{Fe}/\mathrm{H}] \approx 0.29$ for WASP-189 \citep{Lendl2020} and interpolating $\Lambda_\odot(T_\mathrm{ps})$ from solar-metallicity cooling curves \citep{Schure2009,Wang2014}. The shock crossing time is $\tau_{\rm cross} = \sigma_{\rm sh}/v_{\rm sh}$.

Table~\ref{tab:cooling_times} lists the resulting $\tau_{\rm cool}$ and $\tau_{\rm cross}$ for three stellar wind cases. In Cases~2 and 3, where $v_{\rm sw}$ is high and the bow shock is detached, $\tau_{\rm cool} \gtrsim \tau_{\rm cross}$, implying that shocked material is advected past the stand-off region before significant cooling and recombination occur. In these situations, the column of neutral or low-ionisation species setting the NUV opacity is theoretically expected to remain too small to be detectable, corresponding to the absence of an EI signature.

\begin{table}[h!]
\caption{Shock cooling and crossing time calculations for three stellar wind cases.}\label{tab:cooling_times}
\centering
\small
\resizebox{\columnwidth}{!}{%
\begin{tabular}{lcccc}
\hline\hline
Parameter & Unit & Case 1 & Case 2 & Case 3 \\
\hline
$v_\mathrm{sw}$ & km~s$^{-1}$ & 572.97 & 1145.94 & 1718.91 \\
$v_\mathrm{sh}$ & km~s$^{-1}$ & 603.29 & 1161.4 & 1729.25 \\
$\log T_\mathrm{ps}$ & K & 6.86 & 7.43 & 7.78 \\
$\log\Lambda_\odot(T_\mathrm{ps})$ & erg~cm$^3$~s$^{-1}$ & $-22.3329$ & $-22.6212$ & $-22.5360$ \\
$\log\Lambda(T_\mathrm{ps})$ & erg~cm$^3$~s$^{-1}$ & $-22.0429$ & $-22.3312$ & $-22.2461$ \\
$\tau_\mathrm{cool}$ & s & 90.14 & 650.38 & 1196.89 \\
$\tau_\mathrm{cross}$, $\sigma_\mathrm{sh}$$=$$2.45\,R_\mathrm{p}$ & s & 470.05 & 244.17 & 163.99 \\
$\tau_\mathrm{cross}$, $\sigma_\mathrm{sh}$$=$$2.94\,R_\mathrm{p}$ & s & 564.06 & 293.00 & 196.79 \\
\hline
\end{tabular}
}
\tablefoot{
Cooling and crossing times for three stellar wind velocities. $\Lambda_\odot$ and $\Lambda$ are the solar and WASP-189 metallicity-scaled cooling functions. $\tau_\mathrm{cross}$ is calculated for two assumed shock thicknesses $\sigma_\mathrm{sh}$.
}
\end{table}

When the wind slows to trans- or sub-Alfvénic speeds, magnetically guided compression ahead of the obstacle produces a dense pileup even without a strong bow shock, allowing the gas to cool and increasing NUV opacity so that an EI becomes visible. This is most likely during fast-to-slow transitions, when previously shocked gas has not yet dispersed. A-type hosts often have stable, oblique dipolar fields that channel radiatively driven winds into coexisting fast and slow streams, so rotational phase and intrinsic variability naturally shift the system between regimes \citep{Maryline2015}. In general, the interplay between stellar-wind velocity and magnetic confinement, as formulated in the magnetic-confinement paradigm and the Alfvén radius framework, implies that modest variations in wind speed or magnetic-field geometry can shift the system between two regimes: one characterised by a dense, slowly advecting plasma pileup with high NUV opacity, and another by a rapidly advecting, diffuse shock with low NUV opacity \citep{Haemmerle2019}.

\end{appendix}

\end{document}